\documentclass[]{article}

\usepackage{graphicx}
\usepackage[capposition=top]{floatrow}
\usepackage[margin=0.75in]{geometry}
\usepackage{amsmath}

\usepackage{setspace}
\usepackage{enumitem}
\newlist{legal}{enumerate}{10}
\setlist[legal]{label*=\arabic*.}

\title{Machine learning for causal inference: on the use of cross-fit estimators}
\author{Paul N Zivich$^{1, 2}$, Alexander Breskin$^{3}$}
\date{%
	\footnotesize
	$^1$Department of Epidemiology, Gillings School of Global Public Health, University of North Carolina at Chapel Hill, Chapel Hill, NC, USA\\%
	$^2$Carolina Population Center, University of North Carolina at Chapel Hill, Chapel Hill, NC, USA\\%
	$^3$NoviSci, Durham, NC, USA\\[2ex]%
	\today
}

\begin{document}

\maketitle

\begin{abstract}
Modern causal inference methods allow machine learning to be used to weaken parametric modeling assumptions. However, the use of machine learning may result in complications for inference. Doubly-robust cross-fit estimators have been proposed to yield better statistical properties.

We conducted a simulation study to assess the performance of several different estimators for the average causal effect (ACE). The data generating mechanisms for the simulated treatment and outcome included log-transforms, polynomial terms, and discontinuities. We compared singly-robust estimators (g-computation, inverse probability weighting) and doubly-robust estimators (augmented inverse probability weighting, targeted maximum likelihood estimation). Nuisance functions were estimated with parametric models and ensemble machine learning, separately. We further assessed doubly-robust cross-fit estimators.

With correctly specified parametric models, all of the estimators were unbiased and confidence intervals achieved nominal coverage. When used with machine learning, the doubly-robust cross-fit estimators substantially outperformed all of the other estimators in terms of bias, variance, and confidence interval coverage.

Due to the difficulty of properly specifying parametric models in high dimensional data, doubly-robust estimators with ensemble learning and cross-fitting may be the preferred approach for estimation of the ACE in most epidemiologic studies. However, these approaches may require larger sample sizes to avoid finite-sample issues. 

\end{abstract}

\newpage
\section*{INTRODUCTION}

Most causal effect estimation methods for observational data utilize so-called nuisance functions. These functions are not of primary interest but are used as inputs into estimators \cite{Kennedy2016}. For instance, the propensity score function is the nuisance function for the inverse probability of treatment weighted estimator. In low-dimensional settings, these nuisance functions are possible to estimate nonparametrically. However, in more realistic settings, such as those that involve continuous covariates or in which the minimal sufficient adjustment set of confounders is large, parametric models, referred to as nuisance models, are often used to estimate the nuisance functions. Proper specification of these nuisance models is then required for the resulting estimator to be consistent. Specifically, a properly specified model contains the true function as a possible realization. Given the complex underlying relationships between variables, it is often implausible that parametric nuisance models can be properly specified.

Data-adaptive supervised machine learning methods (which are fit using data-driven tuning parameters) \cite{Mooney2018, Bi2019}, have been suggested as an alternative approach to estimate nuisance functions in high-dimensional settings while not imposing overly restrictive parametric functional form assumptions \cite{Schuler2017, Watkins2013, Pirracchio2015, Lee2010, Westreich2010}. Despite this optimism, issues in the use of machine learning for nuisance function estimation have become apparent \cite{Keil2018, Chernozhukov2018, Bahamyirou2019}. Notably, some machine learning algorithms converge to the true answer at a slow rate (i.e., the mean squared error of the estimator diminishes slowly with sample size), leading to substantial under-coverage of corresponding confidence intervals. This slow rate of convergence is the 'cost' of making weaker assumptions. Conversely, the 'cost' of making stronger assumptions is misspecification (i.e., the bias of the estimator does not diminish with sample size at any rate) \cite{Rudolph2018}.

Doubly-robust estimators have several features that make them less prone to model misspecification. First, as implied by their name, these estimators provide two opportunities to properly specify nuisance models. Second, in the context of machine learning methods, doubly-robust estimators allow for the use of slower converging nuisance models. As an added benefit, they permit simple approaches for variance estimation, even when machine-learning is used to fit the nuisance functions. However, restrictions on complexity (i.e. that nuisance models are Donsker) preclude the use of some machine learning approaches. 

Doubly-robust cross-fit estimators have been developed to reduce overfitting and impose less restrictive complexity conditions on the machine learning algorithms used to estimate nuisance functions \cite{Chernozhukov2018, Newey2018}. These estimators relax the convergence conditions to allow for a richer set of algorithms to be used. Cross-fit estimators share similarities with double machine learning \cite{Chernozhukov2018}, cross-validated estimators \cite{Zheng2011, Levy2018}, and sample splitting \cite{Athey2015}. An extension, referred to as double cross-fitting, has recently been proposed to address so-called nonlinearity bias, and when used with undersmoothing methods it achieves the fastest possible convergence rate \cite{Newey2018}. In this work, we detail a general procedure for doubly-robust double cross-fit estimators and demonstrate their performance in a simulation study. We compare a wide range of estimators in the context of a simulated study of statins and subsequent atherosclerotic cardiovascular disease (ASCVD). 

\section*{METHODS}

\subsection*{Data generating mechanism}

Suppose we observe an independent and identically distributed sample $(O_1, ..., O_n)$ where $O_i = (X_i, Y_i, Z_i) \sim F$. Let $X$ indicate statin use; $Y$ indicate the observed incidence of ASCVD; $Y^x$ indicate the potential value $Y$ would have taken if, possibly counter to fact, an individual received treatment $x$; and $Z$ indicate potential confounders. The ACE is then
$$\psi = E[Y^1 - Y^0] = E[Y^1] - E[Y^0]$$
Note that the potential outcomes are not necessarily directly observed, so a set of conditions are needed to identify the ACE from observable data. Specifically, we assume the following conditions hold:
\begin{enumerate}
	\item Counterfactual consistency \cite{Cole2009}
	$$Y_i = Y_i^{x,k} \text{ if }x=X_i, \text{ no matter the value of }k \text{; where }k\text{ is the version of }x$$
	\item Conditional exchangeability \cite{Hernan2006}
	$$Y^x \amalg X|Z \; \forall \; x \in \{0, 1\}$$
	\item Positivity \cite{Westreich2010a}
	$$\text{if } \Pr(Z=z) > 0 \text{ then } \Pr(X=x|Z=z) > 0 \; \forall \; z \in \mathcal{Z}, x \in \{0, 1\} $$
\end{enumerate}
Together, these conditions allow the ACE to be identified as:
$$\int_{z\in\mathcal{Z}} \left(E[Y|X=1, Z=z] - E[Y|X=0, Z=z] \right) dF_Z$$

We considered the following confounders in $Z$ for the simulated data generating process: age ($A$), low-density lipoprotein ($L$), ASCVD risk scores ($R$), and diabetes ($D$). Therefore, $Z=(L,A,R,D)$ in the following simulated data. These factors were chosen based on the 2018 primary prevention guidelines for the management of blood cholesterol \cite{Grundy2019}. Full details on the data generating mechanism are provided in Appendix 1. The incidence of statin use ($X$) was chosen to be similar to reported empirical trends in US adults \cite{Salami2017}, and generated from the following model inspired by the 2018 primary prevention guidelines:
\begin{equation}
\centering
\nonumber	
\begin{split}
	\Pr(X=1|Z) = \text{Bernoulli}(\text{expit}(&-3.471 + 1.390 D_i + 0.112 L_i + 0.973\; I(L_i > \log(60)) \\
	& - 0.046(A_i - 30) + 0.003(A_i - 30)^2 + 0.273\; I(0.05 \le R_i < 0.075)\\
	& + 1.592\; I(0.075 \le R_i < 0.2) + 2.461\; I(R_i \ge 0.2)))
\end{split}
\end{equation}
The ASCVD potential outcomes under each potential value of X were generated from the following model:
\begin{equation}
\centering
\nonumber
\begin{split}
	\Pr(Y^x=1|Z) = \text{Bernoulli}(\text{expit}(&-6.250 - 0.75x + 0.35x(5-L_i) \; I(L_i < \log(130)) + 0.45(A_i - 39)^{0.5}\\ 
	&+ 1.75D_i + 0.29 \exp(R_i + 1) + 0.14 L_i^2 \; I(L_i > \log(120))))
\end{split}
\end{equation}
The observed outcome was calculated as $Y_i = Y_i^1 X_i + (1-X_i) Y_i^0$. The nuisance functions considered are:
\begin{equation}
	\Pr(X=1|Z=z) = \pi(z)
\end{equation}
\begin{equation}
	\Pr(Y|X=x,Z=z) = m_x(z)
\end{equation}
As the name implies, these nuisance functions are not of direct interest but are used for the estimation of the ACE. Unlike in simulated data, the correct specification of these models is often unknown; and in the context of parametric models, must be \textit{a priori} specified. 

\subsection*{Nuisance function estimators}
Before estimating the ACE, the nuisance functions ($\pi(z), m_x(z)$) need to be estimated.  Much of the previous work in causal effect estimation has relied on parametric regression models. However, these models must be sufficiently flexible to capture the true nuisance functions. In our simulations we consider two different parametric model specifications. First, we consider the correct model specification as described previously. This is the best-case scenario for researchers. Unfortunately, this case is unlikely to occur. Second, we considered a main-effects model, where all variables were assumed to be linearly related to the outcome and no interaction terms were included in the model. The main-effects model is quite restrictive and does not contain the true density function.

As an alternative to the parametric models, we consider several data-adaptive machine learning algorithms. There are a variety of potential supervised machine learning algorithms and there is no guarantee on which algorithm will perform best in all scenarios \cite{Wolpert1996}. Therefore, we utilize super-learner with 10-fold cross-validation to estimate the nuisance functions. Super-learner is a generalized ensemble algorithm that allows for the combination of multiple predictive algorithms into a single prediction function \cite{van2007, Rose2013}. Super-learner has been previously shown to asymptotically perform as well as the best performing algorithm included within the super-learner procedure \cite{van2007}, with studies of finite sample performance indicating similar results. Within super-learner, we included the following algorithms: the empirical mean, main-effects logistic regression without regularization, generalized additive model with 4 splines and a ridge penalty of 0.6 \cite{Hastie2017}, generalized additive model with 6 splines, random forest with 500 trees and a minimum of 20 individuals per leaf \cite{Breiman2001}, and a neural network with a single hidden layer consisting of four nodes. Only non-processed main-effects variables were provided to each learner.

\subsection*{Estimators for the ACE}
After estimation of the nuisance functions, the predictions can be used in estimators for the ACE. We considered 4 estimators: g-computation, an inverse probability weighted (IPW) estimator, an augmented inverse probability weighted (AIPW) estimator, and a targeted maximum likelihood estimator (TMLE). The IPW estimator only requires the nuisance function from Equation 1 and g-computation only requires the nuisance function from Equation 2. Due to their reliance on a single nuisance function, these methods are said to be singly-robust. However, these singly-robust estimators require fast convergence of nuisance models, severely limiting which algorithms can be used. The AIPW estimator and TMLE instead use both nuisance functions from both Equations 1 and 2, and have the property of double robustness; such that if either nuisance model is correctly estimated, then the point estimate will be consistent \cite{Schuler2017, Funk2011, Lunceford2004, Bang2005}. Perhaps more important in the context of machine learning, these doubly-robust estimators allow for slower convergence of nuisance models. However, all of these estimators require the nuisance models to not be overly complex, in the sense that they belong to the so-called Donsker class \cite{Chernozhukov2018}. Intuitively, members of this class are less prone to overfitting than models outside the class. For models that do not belong to the Donsker class, confidence intervals may be overly narrow and result in misleading inference. Recent work has demonstrated that cross-fitting paired with doubly-robust estimators weakens the complexity conditions for the nuisance models, which allows for a more diverse set of algorithms. A double cross-fit procedure allows for further theoretical improvements for doubly-robust estimators \cite{Newey2018}. Therefore, we additionally considered double cross-fit alternatives for AIPW (DC-AIPW) and TMLE (DC-TMLE). We briefly outline each estimator, with further details and formulas provided in the Appendix. 

\subsubsection*{G-computation}
We used the g-computation procedure described by Snowden et al. 2011 to estimate the ACE \cite{Snowden2011}. Briefly, the outcome nuisance model, $m_x(z)$, is fit using the observed data. From the fit outcome nuisance model, the probability of $Y$ is predicted under $X=1$ and under $X=0$ for all individuals. The ACE is estimated by taking the average of the differences of the predicted outcome $Y$ under each treatment plan. Wald-type 95\% confidence intervals were generated using the standard deviation of 250 bootstrap samples with replacement, each of size $n$. We note that it is currently unknown whether the bootstrap is generally valid for the g-computation or the other following estimators when data-adaptive methods are used for nuisance model fitting.

\subsubsection*{IPW}
In contrast to g-computation, the IPW estimator relies on estimation of the treatment nuisance model, $\pi(z)$. From the predicted probabilities of $X$, weights are constructed by taking the inverse of the predicted probabilities of the observed $X$. These weights are then used to calculate the weighted average of $Y$ among subjects with each value of $X$. We used robust standard errors that ignore the estimation of the nuisance function, which results in conservative variance estimates for the ACE \cite{Hernan2002}. Therefore, confidence interval coverage is expected to be at least 95\% when the nuisance model is properly specified.

\subsubsection*{AIPW}
The AIPW estimator uses both the treatment and outcome nuisance functions predictions to estimate the ACE. Predicted probabilities of the treatment and outcome are combined via a single equation to generate predictions under each value of $X$, with confidence intervals calculated from the influence curve \cite{Funk2011, Lunceford2004}. 

\subsubsection*{TMLE}
TMLE similarly uses both the treatment and outcome nuisance functions to construct a single estimate. Unlike the AIPW estimator, TMLE uses a 'targeting' step that corrects the bias-variance tradeoff in the estimation \cite{Schuler2017}. This is accomplished by fitting a parametric working model, where the observed $Y$ is modeled as a function of a transformation of the predicted probabilities of $X$ (often referred to as the ‘clever covariate’) with the outcome nuisance model predictions included as an offset. The targeted predictions under each value of $X$ from this model are averaged, and their difference provides an estimate of the ACE. Confidence intervals are calculated from the influence curve. 

\subsubsection*{Double cross-fit}
\begin{figure}
	\centering
	\caption {General double cross-fit procedure for doubly-robust estimators}
	\includegraphics[width=0.9\linewidth]{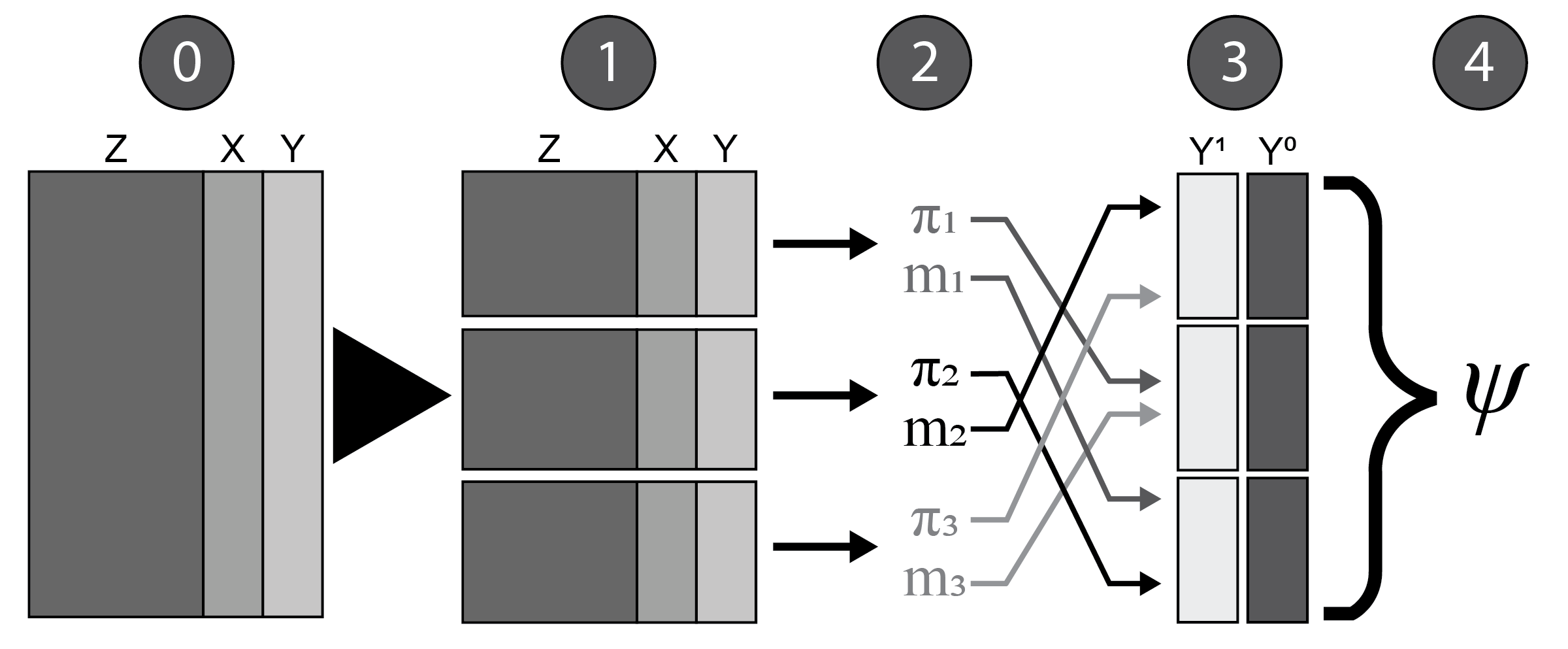}
	\label{fig1}
	\floatfoot{Step 0) The exposure ($X$), outcome ($Y$), and minimally sufficient adjustment set for identification ($Z$) are selected and collected.\\
	Step 1) The data is partitioned into three approximately equal-sized sample splits.\\
	Step 2) The treatment nuisance model and the outcome nuisance model are fit in each sample split.\\
	Step 3) Predicted outcomes under each treatment are estimated using the nuisance models estimated using discordant data sets. For example, sample split 1 uses the treatment nuisance model from sample split 3 and the outcome nuisance model from sample split 2. \\
	Step 4) The target parameter is calculated from the mean of the predictions across all splits. The variance for that particular sample split is calculated as the mean of variances for each split. \\
	Steps 1-4 are repeated a number of times to reduce sensitivity to particular sample splits. The overall point estimate is calculated as the median of the point estimates for all of the different splits. The estimated variance consists of two parts: the variability of the ACE within a particular split and the variance of the ACE point estimate between each split.}
\end{figure}

A visualization of the double cross-fit procedure is provided in Figure \ref{fig1}. This process is compatible with both doubly-robust estimators previously described. First, the data set is randomly partitioned into three approximately equal-sized splits or groups (although this can be generalized to numbers larger than three \cite{Chernozhukov2018, Newey2018}). Note that the splits are non-overlapping, so that each subject belongs to a single split. Second, nuisance models for the treatment and outcome nuisance functions are estimated in each of the three sample splits. This involves using the super-learner fitting procedure independently for each split (for a total of six times - three for the outcome model and three for the treatment model). Third, predicted treatment probabilities and expected values for outcomes are calculated from the nuisance models in the discordant splits, such that the predictions from super-learner do not come from the same data used to fit the models. For example, sample split 1 could have the probability of treatment predicted with the treatment nuisance model from split 3; and the expected value of the outcome predicted with the outcome nuisance model from split 2. The doubly-robust estimator of choice is used to estimate the ACE from the treatment and outcome predictions within each split. For the AIPW estimator this consists of calculating the ACE via the equation provided in Appendix 2. For TMLE this consists of the targeting step. In a final step, the split-specific ACE estimates from all splits are averaged together to produce a final point estimate of the partition-specific ACE.

Since the ACE is dependent on a particular partition of the input data, the previous procedure is repeated a large number of times with different possible partitions. In our implementation, we used 100 different splits as is recommended in other cross-fitting procedures \cite{Chernozhukov2018}. Results were unstable when only using few partitions with flexible algorithms (see Appendix 4). The overall point estimate of the ACE is calculated as the median of the ACE for all partitions $p$
$$\widetilde{ACE} = \text{median}(ACE_p)$$
While the mean can also be chosen, it is more susceptible to outliers and may require a larger number of different partitions \cite{Chernozhukov2018}. The estimated variance for the ACE consists of two parts: the variability of the ACE within a particular partition and the variance of the ACE point estimate between each partition. The variance for the $p$ different splits, is the median of
$$Var(\widetilde{ACE}) = \text{median}(Var(ACE_p) + (ACE_p - \widetilde{ACE})^2)$$
A detailed description of the double cross-fit procedure is provided in Appendix 3.

\subsection*{Performance metrics}
For each combination of nuisance function estimators and ACE estimators; we calculated bias, empirical standard error (ESE), root mean squared error (RMSE), average standard error (ASE), average confidence limit difference (CLD), and 95\% confidence limit coverage over 2000 simulated samples. Bias was defined as mean of the estimated ACE from each simulation minus the true population ACE. ESE was calculated as the standard error of all estimates for an estimator. RMSE was defined the square root of bias squared plus ESE squared. ASE was the mean of the estimated standard error from each simulation. CLD was the mean of the upper confidence limit minus the lower confidence limit. 95\% confidence interval coverage was calculated as the proportion of confidence intervals containing the true population ACE.

All simulations were conducted using Python 3.5.1 (Python Software Foundation, Wilmington, DE) with the sklearn implementations of the previously described algorithms \cite{Pedregosa2011}. Simulation code is available at \\https://github.com/pzivich/publications-code. Outside of the specified parameters, the defaults of the software were used. The true ACE was calculated as the average difference in potential outcomes for a population of 10,000,000 individuals. Simulations for estimation of the ACE were repeated 2000 times for n=3000. The sample size for simulations was chosen such that when split into three equally sized groups (n=1000), the true parametric models could be fit and used to correctly estimate the true ACE.

\section*{RESULTS}
Before presentation of the full simulation results, we present summary statistics and estimates for a single simulated data set. Characteristics of the single study sample are displayed in Table \ref{table1}. Results for the estimators are presented in Table \ref{table2}. Nuisance models estimated with machine learning led to substantially narrower confidence intervals as indicated by the CLD. Differences were less stark for the double cross-fit estimators. Broadly, run-times for estimators of ACE were short. The double cross-fit estimators had substantially longer run times due to the repeated sample splitting procedure. As reference, a single estimation of the DC-AIPW estimator required 600 different super-learner procedures to be fit. There was notable variation between estimates from the different partitions for DC-AIPW with parametric (interquartile range (IQR): -0.10 – -0.08; Range: -0.12 – -0.07) and machine learning (IQR: -0.12 – -0.11; Range: -0.15 – -0.08) nuisance models. A similar pattern was observed with DC-TMLE for parametric (IQR: -0.13 – -0.12; Range: -0.14 – -0.09) and machine learning (IQR: -0.11 – -0.11; Range: -0.12 – -0.07)) as well.

\begin{table}[]
	\centering
	\caption{Descriptive characteristics for a single sample}
	\label{table1}
	\begin{tabular}{lcc}
		\hline
		& Statin (n=776) & No statin (n=2224) \\ \cline{2-3} 
		Age (SD)        & 58 (9.5)       & 53 (7.6)           \\
		Diabetes        & 31\%           & 1\%                \\
		log(LDL) (SD)   & 4.92 (0.2)     & 4.86 (0.2)         \\
		Risk score (SD) & 0.15 (0.2)     & 0.06 (0.1)         \\
		ASCVD           & 37\%           & 29\%               \\ \hline
	\end{tabular}
	\floatfoot{Descriptive statistics for a single sample from mthe data generating mechanism. Continuous variables are preseted as mean (standard deviation)\\
	SD: standard deviation, LDL: low-density lipoproteins, ASCVD: atherosclerotic cardiovascular disease.
	}
\end{table}

\begin{table}[]
	\centering
	\caption{Estimated risk differences for a single sample from the data generating mechanism with main-terms}
	\label{table2}
	\begin{tabular}{llccccc}
		\hline
		&                       & RD    & SD(RD) & 95\% CL      & CLD  & Run-time* \\ \cline{3-7} 
		\multicolumn{2}{l}{G-computation} &       &        &              &      &           \\
		& Main-effects          & -0.14 & 0.016  & -0.17, -0.11 & 0.06 & 0.9       \\
		& Machine learning      & -0.09 & 0.015  & -0.12, -0.06 & 0.06 & 82.3      \\
		\multicolumn{2}{l}{IPW}      &       &        &              &      &           \\
		& Main-effects          & -0.13 & 0.039  & -0.20, -0.05 & 0.15 & 0.0       \\
		& Machine learning      & -0.11 & 0.028  & -0.16, -0.05 & 0.11 & 0.3       \\
		\multicolumn{2}{l}{AIPW}      &       &        &              &      &           \\
		& Main-effects          & -0.08 & 0.038  & -0.16, -0.01 & 0.15 & 0.0       \\
		& Machine learning      & -0.11 & 0.016  & -0.14, -0.08 & 0.06 & 0.7       \\
		\multicolumn{2}{l}{TMLE}      &       &        &              &      &           \\
		& Main-effects          & -0.12 & 0.029  & -0.18, -0.06 & 0.11 & 0.0       \\
		& Machine learning      & -0.12 & 0.016  & -0.15, -0.09 & 0.06 & 0.7       \\
		\multicolumn{2}{l}{DC-AIPW}   &       &        &              &      &           \\
		& Main-effects          & -0.09 & 0.039  & -0.16, -0.01 & 0.15 & 1.3       \\
		& Machine learning      & -0.11 & 0.023  & -0.16, -0.07 & 0.09 & 128.1     \\
		\multicolumn{2}{l}{DC-TMLE}   &       &        &              &      &           \\
		& Main-effects          & -0.12 & 0.029  & -0.18, -0.07 & 0.11 & 1.3       \\
		& Machine learning      & -0.11 & 0.021  & -0.15, -0.07 & 0.08 & 129.9     \\ \hline
	\end{tabular}
	\floatfoot{RD: risk difference, SD(RD): standard deviation for the risk difference, 95\% CL: 95\% confidence limits, CLD: confidence limit difference defined as the upper confidence limit minus the lower confidence limit.\\
	Machine learning estimators were super-learner with 10-fold cross validation. Algorithms included were the empirical mean, main-effects logistic regression without regularization, generalized additive model with 4 splines and a ridge penalty of 0.6, generalized additive model with 6 splines, random forest with 500 trees and a minimum of 20 individuals per leaf, and a neural network with a single hidden layer consisting of four nodes.\\
	Double cross-fit procedures included 100 different sample splits.\\
	* Run times are based on a server running on a single 2.5 GHz processor with 5 GB of memory allotted. Run times are indicated in minutes. G-computation run-times are large due to the use of a bootstrap procedure to calculate the variance for the risk difference. IPW used robust variance estimators. AIPW, TMLE, DC-AIPW, and DC-TMLE variances were calculated using influence curves.
	}
\end{table}

\subsection*{Simulations}
\begin{figure}
	\centering
	\caption {Bias and confidence interval coverage of estimators of the average causal effect}
	\includegraphics[width=0.9\linewidth]{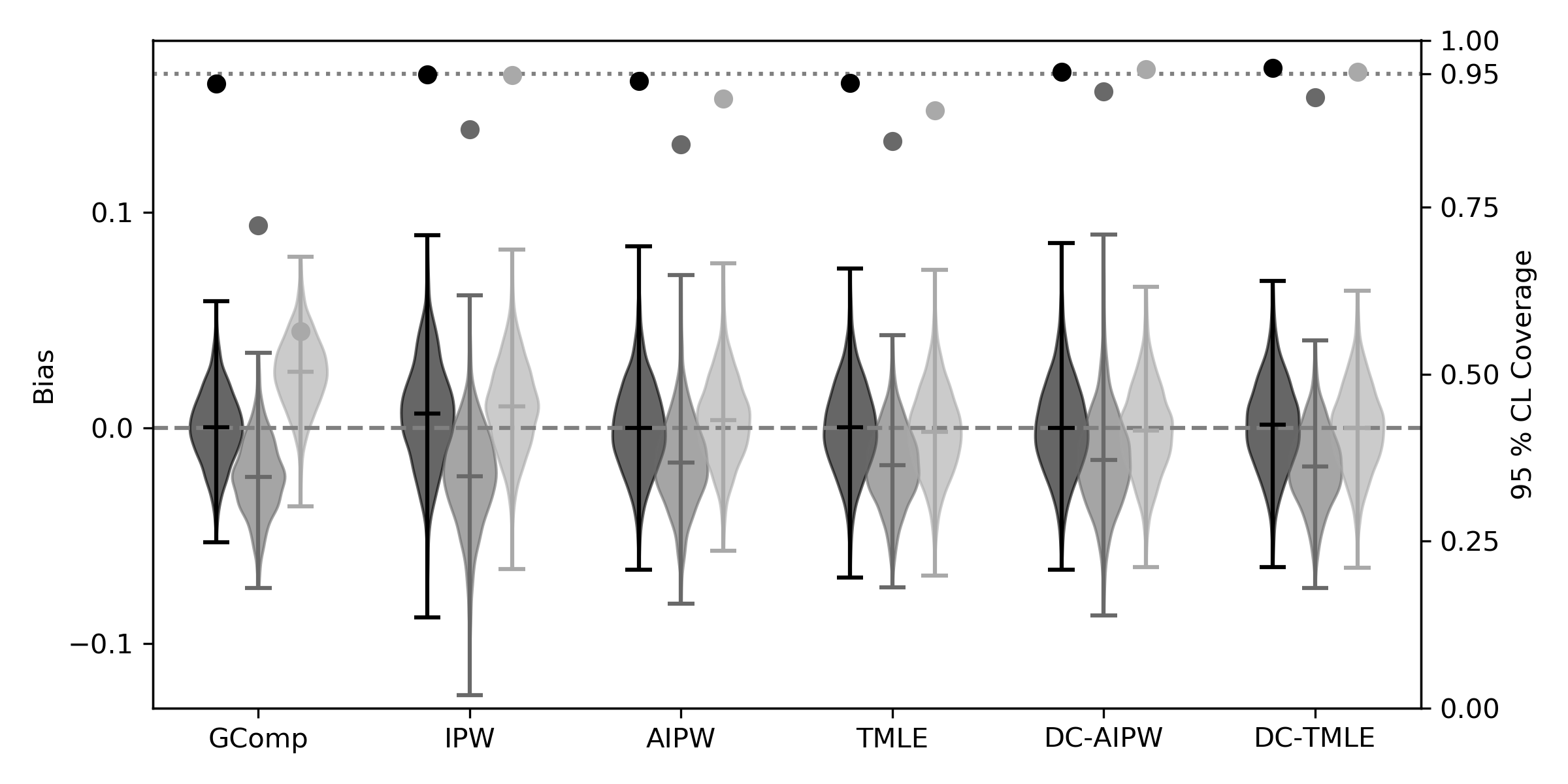}
	\label{fig2}
	\floatfoot{GComp: g-computation, IPW: inverse probability of treatment weighted estimator, AIPW: augmented inverse probability weighted estimator, TMLE: targeted maximum likelihood, DC-AIPW: double cross-fit AIPW, DC-TMLE: double cross-fit TMLE}
\end{figure}

As expected, ACE estimators with correctly specified parametric nuisance models were unbiased and confidence intervals resulted in near-95\% coverage (Figure \ref{fig2}, Table \ref{table3}). The most efficient estimator was g-computation (ESE=0.017), followed by TMLE (ESE=0.021), AIPW (ESE=0.021), and IPW (ESE=0.023). DC-TMLE and DC-AIPW were comparable to their non cross-fit counterparts (0.021 and 0.021, respectively). Confidence interval coverage was higher for double cross-fit estimators.

For main-effects parametric nuisance models, all ACE estimators were biased from the true target parameter. Increased RMSE was primarily a result of the occurrence of bias. The double cross-fit procedure did not improve estimates in terms of bias due to model misspecification. Confidence interval coverage was likely greater solely due to the penalty in estimated variance due to variation between partitions. 

For singly-robust estimators with machine learning, bias increased compared to correctly specified parametric models (Table \ref{table3}, Figure \ref{fig2}). Non-cross-fit doubly robust estimators with machine learning resulted in unbiased estimates of the ACE, but confidence interval coverage was below expected levels for AIPW (91.1\%) and TMLE (89.5\%). Confidence interval coverage of DC-AIPW and DC-TMLE were near nominal levels (95.6\% and 95.0\%, respectively). 

\begin{table}[]
	\centering
	\caption{Simulation results for estimators under different approaches to estimation of the nuisance functions}
	\label{table3}
	\begin{tabular}{llcccccc}
		\hline
		&                       & Bias   & RMSE  & ASE   & ESE   & CLD   & Coverage \\ \cline{3-8} 
		\multicolumn{2}{l}{G-computation} &        &       &       &       &       &          \\
		& True                  & 0.000  & 0.017 & 0.017 & 0.017 & 0.065 & 93.5\%   \\
		& Main-effects          & -0.023 & 0.029 & 0.017 & 0.018 & 0.067 & 72.3\%   \\
		& Machine learning      & 0.026  & 0.031 & 0.015 & 0.017 & 0.058 & 56.5\%   \\
		\multicolumn{2}{l}{IPW}      &        &       &       &       &       &          \\
		& True                  & 0.007  & 0.025 & 0.025 & 0.024 & 0.097 & 94.9\%   \\
		& Main-effects          & -0.022 & 0.032 & 0.023 & 0.023 & 0.091 & 86.6\%   \\
		& Machine learning      & 0.010  & 0.023 & 0.023 & 0.021 & 0.090 & 94.8\%   \\
		\multicolumn{2}{l}{AIPW}      &        &       &       &       &       &          \\
		& True                  & 0.000  & 0.021 & 0.020 & 0.021 & 0.077 & 93.9\%   \\
		& Main-effects          & -0.016 & 0.026 & 0.020 & 0.020 & 0.076 & 84.4\%   \\
		& Machine learning      & 0.004  & 0.020 & 0.017 & 0.019 & 0.066 & 91.3\%   \\
		\multicolumn{2}{l}{TMLE}      &        &       &       &       &       &          \\
		& True                  & 0.000  & 0.021 & 0.020 & 0.021 & 0.077 & 93.6\%   \\
		& Main-effects          & -0.017 & 0.025 & 0.019 & 0.018 & 0.075 & 84.9\%   \\
		& Machine learning      & -0.002 & 0.020 & 0.017 & 0.020 & 0.065 & 89.5\%   \\
		\multicolumn{2}{l}{DC-AIPW}   &        &       &       &       &       &          \\
		& True                  & 0.000  & 0.021 & 0.022 & 0.021 & 0.085 & 95.2\%   \\
		& Main-effects          & -0.015 & 0.026 & 0.027 & 0.022 & 0.106 & 92.4\%   \\
		& Machine learning      & -0.001 & 0.020 & 0.021 & 0.020 & 0.082 & 95.6\%   \\
		\multicolumn{2}{l}{DC-TMLE}   &        &       &       &       &       &          \\
		& True                  & 0.001  & 0.020 & 0.021 & 0.020 & 0.084 & 95.8\%   \\
		& Main-effects          & -0.018 & 0.025 & 0.024 & 0.018 & 0.094 & 91.4\%   \\
		& Machine learning      & 0.000  & 0.020 & 0.020 & 0.020 & 0.079 & 95.2\%   \\ \hline
	\end{tabular}
	\floatfoot{RMSE: root mean squared error, ASE: average standard error, ESE: empirical standard error, CLD: confidence limit difference, Coverage: 95\% confidence limit coverage of the true value.\\
	IPW: inverse probability of treatment weights, AIPW: augmented inverse probability of treatment weights, TMLE: targeted maximum likelihood estimator, DC-AIPW: double cross-fit AIPW, DC-TMLE: double cross-fit TMLE. \\
	True: correct model specification. Main-effects: all variables were assumed to be linearly related to the outcome and no interaction terms were included in the model. Machine learning: super-learner with 10-fold cross-validation including empirical mean, main-effects logistic regression without regularization, generalized additive models, random forest, and a neural network.
	}
\end{table}

\section*{DISCUSSION}
In the unlikely scenario in which parametric nuisance model specifications correctly capture the true function, all estimators considered are consistent and subsequent inference is valid. Confidence intervals were wider for double cross-fit estimators due to the variance between partitions being incorporated from the sample splitting procedure. The increase in CLD highlights the bias-precision trade-off made when choosing less-restrictive estimators. However, it is often unreasonable to assume correct parametric model specification in high-dimensional data with weak background information or theory. The pursuit of weaker parametric assumptions for nuisance model specification is worthwhile, with machine learning being a viable approach. However, naive use of machine learning may lead to bias and incorrect inference. As highlighted in our simulation, doubly robust estimators with double cross-fitting and machine learning outperformed both estimators with incorrectly specified parametric nuisance models and non-cross-fit estimators with machine learning. While the bias of the IPW estimator fit with machine learning was small compared with g-computation and confidence interval coverage achieved a nominal level, the variance was substantially larger than any other estimator; highlighting the inefficiency of this method. Further, there is currently no theory supporting valid statistical inference for singly-robust estimators with machine learning \cite{Kennedy2017a, Naimi2017, Vaart2014}. In summation, doubly-robust estimators with machine learning and cross-fitting may be the preferred approach for ACE estimation in many epidemiologic studies.

The need for doubly-robust estimators with cross-fitting when using data-adaptive machine learning for nuisance function estimation arises from two terms in the Von Mises expansion of the estimator \cite{Kennedy2016}. The first term, which is described by an empirical process term in the expansion, can be controlled by either restricting the complexity of the nuisance models (e.g., by requiring them to be in the Donsker class) or through cross-fitting. Because it can be difficult or impossible to verify that a given machine learning method is in the Donsker class, cross-fitting provides a simple and attractive alternative. The second term is the second-order remainder, and it converges to zero as the sample size increases. For valid inference, it is desirable for this remainder term to converge as a function of $n^{-1/2}$, referred to as as ‘root-n’ convergence. Convergence rates are not a computational issue, but rather a feature of the estimator itself. Unfortunately, data-adaptive algorithms often have slower convergence rates as a result of their flexibility. However, because the second-order remainder term of doubly-robust estimators is the product of the approximation errors of the treatment and outcome nuisance models, doubly-robust estimators only require that \textit{the product of the convergence rates} for nuisance models be $n^{-1/2}$. To summarize, cross-fitting permits the use of highly complex nuisance models, while doubly-robust estimators permit the use of slowly converging nuisance models. Used together, these approaches allow one to use a wide class of data-adaptive machine learning methods to estimate causal effects.

Cross-fitting has had a long history in statistics \cite{Bickel1988, Pfanzagl1990, Hajek1962}, and recent emphasis has focused on its use for nonparametric nuisance function estimation \cite{Chernozhukov2018, Newey2018, Robins2008, Bickel1982}. Broadly, cross-fit procedures can be seen as an approach to avoid the overfitting of nuisance models. Single cross-fit procedures, where both nuisance models are fit in a single split and predictions are made in a second split, uncouple the nuisance model estimation from the corresponding predicted values, preventing so-called own observation bias \cite{Newey2018}. However, the treatment nuisance model and outcome nuisance model are estimated using the same data in single cross-fit procedures. Double cross-fit procedures decouple these nuisance models by using separate splits, removing so-called nonlinearity bias \cite{Newey2018}. Removing this secondary bias term may further improve the performance of doubly-robust estimators. When certain undersmoothing methods are additionally used, the double cross-fit procedure achieves the fastest known convergence rate of any estimator. As demonstrated in the simulations, even if these undersmoothing methods are not used double cross-fitting results in tangible benefits regarding point estimation and inference with machine learning algorithms. 

While cross-fitting has tangible benefits, these benefits are not without cost. First, run-times for the double cross-fit estimators are substantially longer due to the repetition of fitting algorithms to a variety of different partitions. We note that the double cross-fit procedure can easily be made to run in parallel, substantially reducing run-times. Computational costs may limit cross-fit procedures to estimators with closed-form variances, since bootstrapping would require considerable computational resources. A second, and potentially more problematic, cost is that sample splitting procedures reduce the amount of data available with which to fit algorithms. While the asymptotic behavior of the estimator is the same as if the entire sample had been used (indeed each data point contributes to both nuisance function and parameter estimation), the partitioning of finite data may preclude some complex algorithms from use. This finite data problem is exacerbated with the use of k-fold super-learner, further stretching the available data to each model fit. For data sets with few observations, increasing the number of folds in super-learner may aid in alleviating this issue \cite{Naimi2018}. The use of single cross-fit at minimum procedures may also aid with finite sample issues since, instead of partitioning the data into three splits, a single cross-fit requires partitioning the data in half. However, the flexibility of machine learning for nuisance function estimation may be limited in these small data sets to begin with \cite{Keil2018}. Whether single cross-fit with machine learning or highly flexible parametric models is preferred in these scenarios is an area for future study.

The problems of sample splitting can manifest themselves as random violations of the positivity assumption \cite{Petersen2012}. As detailed in previous work by Yu et al. 2019, confounders that are strongly related to the exposure may result in positivity violations \cite{Yu2019}. Due to the flexibility of machine learning algorithms, these positivity issues may result in highly variable estimates. Furthermore, positivity issues may not be easy to diagnose, especially in procedures like double cross-fitting. Similar to previous recommendations \cite{Yu2019}, using multiple approaches to triangulate estimates may be helpful. For example, researchers may want to compare a flexible parametric AIPW estimator and a cross-fit AIPW estimator with super-learner.

While our results support the use of machine learning algorithms, machine learning is not a panacea for causal inference. Rather, machine learning can be seen as weakening a single assumption, namely the assumption of proper model specification. Prior substantive knowledge to justify counterfactual consistency, conditional exchangeability, and positivity remain necessary for causal inference \cite{Keil2018, Naimi2018}. For super learner and other ensemble approaches to provide the maximal benefit in terms of specification, a diverse set of algorithms should be included \cite{Rose2013}. Furthermore, multiple tuning parameters, sometimes referred to as hyperparameters, should be included. While the program defaults are often used, these hyperparameters can dramatically change performance of algorithms \cite{Dominici2002}. Therefore, super learner should not only include a diverse set of algorithms, but also those same algorithms under a diverse set of hyperparameters. Our simulations did not extensively explore hyperparameters; with the inclusion of only two hyperparameter specifications for generalized additive models. Because double cross-fit procedures scale poorly in terms of run-time with the addition of algorithms, including more algorithms with different hyperparameters can have substantial cost in terms of run-time. Depending on the complexity of machine learning algorithms used, alternative approaches may be required for hyperparameter tuning within the cross-fitting procedure \cite{Wong2019}. Despite these concerns, a wide variety of hyperparameters should be explored in applications of double cross-fitting. Lastly, variable transformations (e.g. interaction terms, etc.) may be necessary for performance and should be done in practice \cite{Naimi2017}. 

Future work is needed to compare the performance of single cross-fit, double cross-fit, and other alternatives, such as cross-validated TMLE \cite{Levy2018}, to weaken nuisance model restrictions under a variety of data generating mechanisms. Other work is needed to develop diagnostics for cross-fitting and to potentially allow the addition of other nuisance functions. Due to the repetition of partitioning, standard diagnostics (e.g. examining the distributions of predicted treatment probabilities \cite{Yu2019}) may be more difficult to interpret. Additionally, realistic analyses often have additional issues that must be addressed, such as missing data and loss-to-follow-up. Therefore, additional nuisance functions (like inverse probability weights for informative loss-to-follow-up) are often needed and cross-fit procedures for these scenarios should be assessed.

\subsection*{Conclusions}
Machine learning is not a magic formula for the monumental task of causal effect estimation. However, these algorithms do impose less restrictive assumptions regarding the possible forms of the nuisance functions used for estimation. Cross-fit estimators should be seen as an approach to allow for flexibly estimating nuisance functions while retaining valid inference. In practice, cross-fit estimators should be used regularly with a super-learner that includes a diverse library of learners.

\section*{Acknowledgments}
The authors would like to thank Ashley Naimi, Edward Kennedy, and Stephen Cole for their advice and discussion; and the three anonymous peer reviewers whose comments helped improve the clarity of this manuscript. We would like to further thank the University of North Carolina at Chapel Hill and the Research Computing group for providing computational resources that have contributed to these results.
PNZ received training support (T32-HD091058, PI: Aiello, Hummer) from the National Institutes of Health.

\bibliography{biblio}{}

\begin{thebibliography}{10}

\bibitem{Kennedy2016}
E.~H. Kennedy, ``Semiparametric theory and empirical processes in causal
  inference,'' in {\em Statistical causal inferences and their applications in
  public health research}, pp.~141--167, Springer, 2016.

\bibitem{Mooney2018}
S.~J. Mooney and V.~Pejaver, ``Big data in public health: terminology, machine
  learning, and privacy,'' {\em Annual review of public health}, vol.~39,
  pp.~95--112, 2018.

\bibitem{Bi2019}
Q.~Bi, K.~E. Goodman, J.~Kaminsky, and J.~Lessler, ``What is machine learning?
  a primer for the epidemiologist,'' {\em American journal of epidemiology},
  2019.

\bibitem{Schuler2017}
M.~S. Schuler and S.~Rose, ``Targeted maximum likelihood estimation for causal
  inference in observational studies,'' {\em American journal of epidemiology},
  vol.~185, no.~1, pp.~65--73, 2017.

\bibitem{Watkins2013}
S.~Watkins, M.~Jonsson-Funk, M.~A. Brookhart, S.~A. Rosenberg, T.~M. O'Shea,
  and J.~Daniels, ``An empirical comparison of tree-based methods for
  propensity score estimation,'' {\em Health services research}, vol.~48,
  no.~5, pp.~1798--1817, 2013.

\bibitem{Pirracchio2015}
R.~Pirracchio, M.~L. Petersen, and M.~van~der Laan, ``Improving propensity
  score estimators' robustness to model misspecification using super learner,''
  {\em American journal of epidemiology}, vol.~181, no.~2, pp.~108--119, 2015.

\bibitem{Lee2010}
B.~K. Lee, J.~Lessler, and E.~A. Stuart, ``Improving propensity score weighting
  using machine learning,'' {\em Statistics in medicine}, vol.~29, no.~3,
  pp.~337--346, 2010.

\bibitem{Westreich2010}
D.~Westreich, J.~Lessler, and M.~J. Funk, ``Propensity score estimation: neural
  networks, support vector machines, decision trees (cart), and
  meta-classifiers as alternatives to logistic regression,'' {\em Journal of
  clinical epidemiology}, vol.~63, no.~8, pp.~826--833, 2010.

\bibitem{Keil2018}
A.~P. Keil and J.~K. Edwards, ``You are smarter than you think:(super) machine
  learning in context,'' {\em European journal of epidemiology}, vol.~33,
  no.~5, pp.~437--440, 2018.

\bibitem{Chernozhukov2018}
V.~Chernozhukov, D.~Chetverikov, M.~Demirer, E.~Duflo, C.~Hansen, W.~Newey, and
  J.~Robins, ``{Double/debiased machine learning for treatment and structural
  parameters},'' {\em The Econometrics Journal}, vol.~21, pp.~C1--C68, 01 2018.

\bibitem{Bahamyirou2019}
A.~Bahamyirou, L.~Blais, A.~Forget, and M.~E. Schnitzer, ``Understanding and
  diagnosing the potential for bias when using machine learning methods with
  doubly robust causal estimators,'' {\em Statistical methods in medical
  research}, vol.~28, no.~6, pp.~1637--1650, 2019.

\bibitem{Rudolph2018}
J.~E. Rudolph, S.~R. Cole, and J.~K. Edwards, ``Parametric assumptions equate
  to hidden observations: comparing the efficiency of nonparametric and
  parametric models for estimating time to aids or death in a cohort of
  hiv-positive women,'' {\em BMC medical research methodology}, vol.~18, no.~1,
  p.~142, 2018.

\bibitem{Newey2018}
W.~K. Newey and J.~R. Robins, ``Cross-fitting and fast remainder rates for
  semiparametric estimation,'' {\em arXiv preprint arXiv:1801.09138}, 2018.

\bibitem{Zheng2011}
W.~Zheng and M.~J. van~der Laan, ``Cross-validated targeted minimum-loss-based
  estimation,'' in {\em Targeted Learning}, pp.~459--474, Springer, 2011.

\bibitem{Levy2018}
J.~Levy, ``An easy implementation of cv-tmle,'' {\em arXiv preprint
  arXiv:1811.04573}, 2018.

\bibitem{Athey2015}
S.~Athey and G.~Imbens, ``Recursive partitioning for heterogeneous causal
  effects,'' {\em arXiv preprint arXiv:1504.01132}, 2015.

\bibitem{Cole2009}
S.~R. Cole and C.~E. Frangakis, ``The consistency statement in causal
  inference: a definition or an assumption?,'' {\em Epidemiology}, vol.~20,
  no.~1, pp.~3--5, 2009.

\bibitem{Hernan2006}
M.~A. Hern{\'a}n and J.~M. Robins, ``Estimating causal effects from
  epidemiological data,'' {\em Journal of Epidemiology \& Community Health},
  vol.~60, no.~7, pp.~578--586, 2006.

\bibitem{Westreich2010a}
D.~Westreich and S.~R. Cole, ``Invited commentary: positivity in practice,''
  {\em American journal of epidemiology}, vol.~171, no.~6, pp.~674--677, 2010.

\bibitem{Grundy2019}
S.~M. Grundy, N.~J. Stone, A.~L. Bailey, C.~Beam, K.~K. Birtcher, R.~S.
  Blumenthal, L.~T. Braun, S.~de~Ferranti, J.~Faiella-Tommasino, D.~E. Forman,
  {\em et~al.}, ``2018 aha/acc/aacvpr/aapa/abc/acpm/ada/ags/apha/aspc/nla/pcna
  guideline on the management of blood cholesterol: a report of the american
  college of cardiology/american heart association task force on clinical
  practice guidelines,'' {\em Journal of the American College of Cardiology},
  vol.~73, no.~24, pp.~e285--e350, 2019.

\bibitem{Salami2017}
J.~A. Salami, H.~Warraich, J.~Valero-Elizondo, E.~S. Spatz, N.~R. Desai, J.~S.
  Rana, S.~S. Virani, R.~Blankstein, A.~Khera, M.~J. Blaha, {\em et~al.},
  ``National trends in statin use and expenditures in the us adult population
  from 2002 to 2013: insights from the medical expenditure panel survey,'' {\em
  JAMA cardiology}, vol.~2, no.~1, pp.~56--65, 2017.

\bibitem{Wolpert1996}
D.~H. Wolpert, ``The lack of a priori distinctions between learning
  algorithms,'' {\em Neural computation}, vol.~8, no.~7, pp.~1341--1390, 1996.

\bibitem{van2007}
M.~J. Van~der Laan, E.~C. Polley, and A.~E. Hubbard, ``Super learner,'' {\em
  Statistical applications in genetics and molecular biology}, vol.~6, no.~1,
  2007.

\bibitem{Rose2013}
S.~Rose, ``Mortality risk score prediction in an elderly population using
  machine learning,'' {\em American journal of epidemiology}, vol.~177, no.~5,
  pp.~443--452, 2013.

\bibitem{Hastie2017}
T.~J. Hastie, ``Generalized additive models,'' in {\em Statistical models in
  S}, pp.~249--307, Routledge, 2017.

\bibitem{Breiman2001}
L.~Breiman, ``Random forests,'' {\em Machine learning}, vol.~45, no.~1,
  pp.~5--32, 2001.

\bibitem{Funk2011}
M.~J. Funk, D.~Westreich, C.~Wiesen, T.~St{\"u}rmer, M.~A. Brookhart, and
  M.~Davidian, ``Doubly robust estimation of causal effects,'' {\em American
  journal of epidemiology}, vol.~173, no.~7, pp.~761--767, 2011.

\bibitem{Lunceford2004}
J.~K. Lunceford and M.~Davidian, ``Stratification and weighting via the
  propensity score in estimation of causal treatment effects: a comparative
  study,'' {\em Statistics in medicine}, vol.~23, no.~19, pp.~2937--2960, 2004.

\bibitem{Bang2005}
H.~Bang and J.~M. Robins, ``Doubly robust estimation in missing data and causal
  inference models,'' {\em Biometrics}, vol.~61, no.~4, pp.~962--973, 2005.

\bibitem{Snowden2011}
J.~M. Snowden, S.~Rose, and K.~M. Mortimer, ``Implementation of g-computation
  on a simulated data set: demonstration of a causal inference technique,''
  {\em American journal of epidemiology}, vol.~173, no.~7, pp.~731--738, 2011.

\bibitem{Hernan2002}
M.~A. Hern{\'a}n, B.~A. Brumback, and J.~M. Robins, ``Estimating the causal
  effect of zidovudine on cd4 count with a marginal structural model for
  repeated measures,'' {\em Statistics in medicine}, vol.~21, no.~12,
  pp.~1689--1709, 2002.

\bibitem{Pedregosa2011}
F.~Pedregosa, G.~Varoquaux, A.~Gramfort, V.~Michel, B.~Thirion, O.~Grisel,
  M.~Blondel, P.~Prettenhofer, R.~Weiss, V.~Dubourg, {\em et~al.},
  ``Scikit-learn: Machine learning in python,'' {\em Journal of machine
  learning research}, vol.~12, no.~Oct, pp.~2825--2830, 2011.

\bibitem{Kennedy2017a}
E.~H. Kennedy and S.~Balakrishnan, ``Discussion of" data-driven confounder
  selection via markov and bayesian networks" by jenny h{\"a}ggstr{\"o}m,''
  {\em arXiv preprint arXiv:1710.11566}, 2017.

\bibitem{Naimi2017}
A.~I. Naimi and E.~H. Kennedy, ``Nonparametric double robustness,'' {\em arXiv
  preprint arXiv:1711.07137}, 2017.

\bibitem{Vaart2014}
A.~van~der Vaart, ``Higher order tangent spaces and influence functions,'' {\em
  Statistical Science}, pp.~679--686, 2014.

\bibitem{Bickel1988}
P.~J. Bickel and Y.~Ritov, ``Estimating integrated squared density derivatives:
  sharp best order of convergence estimates,'' {\em Sankhy{\=a}: The Indian
  Journal of Statistics, Series A}, pp.~381--393, 1988.

\bibitem{Pfanzagl1990}
J.~Pfanzagl, ``Estimation in semiparametric models,'' in {\em Estimation in
  Semiparametric Models}, pp.~17--22, Springer, 1990.

\bibitem{Hajek1962}
J.~H{\'a}jek, ``Asymptotically most powerful rank-order tests,'' {\em The
  Annals of Mathematical Statistics}, pp.~1124--1147, 1962.

\bibitem{Robins2008}
J.~Robins, L.~Li, E.~Tchetgen, A.~van~der Vaart, {\em et~al.}, ``Higher order
  influence functions and minimax estimation of nonlinear functionals,'' in
  {\em Probability and statistics: essays in honor of David A. Freedman},
  pp.~335--421, Institute of Mathematical Statistics, 2008.

\bibitem{Bickel1982}
P.~J. Bickel, ``On adaptive estimation,'' {\em The Annals of Statistics},
  pp.~647--671, 1982.

\bibitem{Naimi2018}
A.~I. Naimi and L.~B. Balzer, ``Stacked generalization: an introduction to
  super learning,'' {\em European journal of epidemiology}, vol.~33, no.~5,
  pp.~459--464, 2018.

\bibitem{Petersen2012}
M.~L. Petersen, K.~E. Porter, S.~Gruber, Y.~Wang, and M.~J. Van Der~Laan,
  ``Diagnosing and responding to violations in the positivity assumption,''
  {\em Statistical methods in medical research}, vol.~21, no.~1, pp.~31--54,
  2012.

\bibitem{Yu2019}
Y.-H. Yu, L.~M. Bodnar, M.~M. Brooks, K.~P. Himes, and A.~I. Naimi,
  ``Comparison of parametric and nonparametric estimators for the association
  between incident prepregnancy obesity and stillbirth in a population-based
  cohort study,'' {\em American journal of epidemiology}, vol.~188, no.~7,
  pp.~1328--1336, 2019.

\bibitem{Dominici2002}
F.~Dominici, A.~McDermott, S.~L. Zeger, and J.~M. Samet, ``On the use of
  generalized additive models in time-series studies of air pollution and
  health,'' {\em American journal of epidemiology}, vol.~156, no.~3,
  pp.~193--203, 2002.

\bibitem{Wong2019}
J.~Wong, T.~Manderson, M.~Abrahamowicz, D.~L. Buckeridge, and R.~Tamblyn, ``Can
  hyperparameter tuning improve the performance of a super learner?: A case
  study,'' {\em Epidemiology (Cambridge, Mass.)}, vol.~30, no.~4, p.~521, 2019.

\end{thebibliography}
\bibliographystyle{ieeetr}

\section*{Appendix}

\subsection*{Section 1: Data generating mechanism}
Let $A$ indicate age, $L$ indicate the natural-log transformed low-density lipo-protein, $D$ indicate diabetes, $F$ indicate frailty, and $R$ indicate the risk score. All observations are indexed by $i$, such that $A_i$ refers to the age of individual $i$. All variables were generated from the following distributions in their respective order.
\begin{equation}
\nonumber
A_i = \begin{cases}
75 - \sqrt{30(v_i-60)} & \text{ if } v_i > 60\\
v_i & \text{ if } v_i \le 60 
\end{cases}, \;\;
v = \frac{55\times \text{Uniform}(0, 1) + 80}{2}
\end{equation}
$$L_i = 0.005A_i + \text{Normal}(\log(100), 0.18)$$
$$D_i = \text{Bernoulli}(\text{expit}(-4.23 + 0.03L_i -0.02A_i+0.0009A_i^2))$$
$$F_i = \text{expit}(-5.5 + 0.05(A_i -20) + 0.001A_i^2 +\text{Normal}(0, 1)$$
$$R_i = \text{expit}(4.299 + 3.501D_i - 2.07\log(A_i) +0.051 \log(A_i)^2 +4.090L_i -1.04L_i \log(A_i) +0.01F_i)$$
where $\text{expit}(x) = \frac{\exp(x)}{1 + \exp(x)}$. Statin ($X$) and atherosclerotic cardiovascular disease ($Y$) were generated from the following models
\begin{equation}
	\centering
	\nonumber	
	\begin{split}
	\Pr(X=1|Z) = \text{Bernoulli}(\text{expit}(&-3.471 + 1.390 D_i + 0.112 L_i + 0.973\; I(L_i > \log(60)) \\
	& - 0.046(A_i - 30) + 0.003(A_i - 30)^2 + 0.273\; I(0.05 \le R_i < 0.075)\\
	& + 1.592\; I(0.075 \le R_i < 0.2) + 2.461\; I(R_i \ge 0.2)))
	\end{split}
\end{equation}
\begin{equation}
	\centering
	\nonumber
	\begin{split}
	\Pr(Y^x=1|Z) = \text{Bernoulli}(\text{expit}(&-6.250 - 0.75x + 0.35x(5-L_i) \; I(L_i < \log(130)) + 0.45(A_i - 39)^{0.5}\\ 
	&+ 1.75D_i + 0.29 \exp(R_i + 1) + 0.14 L_i^2 \; I(L_i > \log(120))))
	\end{split}
\end{equation}
From each of the potential outcomes, the observed outcomes were calculated based on the potential outcome under the observed treatment.
$$Y_i = X_i Y_i^1 + (1-X_i)Y_i^0$$

Across all simulations the estimand of interest, the population average causal effect, was defined as
$$\psi = E[Y^1 - Y^0]$$
The true value for $\psi$ was calculated directly from the potential outcomes of 10,000,000 individuals. Let $\hat{\psi}$ indicate the sample-specific estimate of the population average causal effect.

\subsection*{Section 2: Estimators}
We consider two nuisance models, where $m_x(Z_i)$ is the outcome nuisance model and $\pi(Z_i)$ is the treatment nuisance model. The estimated outcome nuisance model is expressed as:
$$\widehat{m_x}(z) = \hat{E}[Y|X=x, Z=z]$$
and the estimated treatment nuisance model is expressed as:
$$\widehat{\pi}(z) = \widehat{\Pr}(X=1|Z=z)$$

\subsubsection*{G-computation}
The g-computation algorithm consists of the outcome nuisance model predictions only. The estimated parameters of the outcome nuisance model are used to predict outcome values. The average of the predicted outcome values under each treatment plan are contrasted. The average causal effect is estimated via:
$$\hat{\psi}_G = n^{-1} \sum_i \widehat{m_1}(Z_i) - n^{-1} \sum_i \widehat{m_0}(Z_i)$$
Wald-type confidence intervals were calculated with the standard error estimated using a bootstrapping procedure. Briefly, $n$ samples are drawn with replacement from the observed sample. Nuisance models are fit to the newly generated sample and used to calculate the average causal effect. This procedure was repeated 250 times, with the bootstrap standard error defined as the standard error of the 250 re-estimated average causal effects. We note that there is no theoretical justification for the use of the bootstrap for singly-robust estimators when data adaptive methods are used to fit the nuisance models.

\subsubsection*{Inverse probability weighted estimator}
The IPW estimator consists of the treatment nuisance model predictions only. The estimated parameters of the treatment nuisance model are used to predict probabilities of treatment. The inverse of these estimated probabilities are used to construct the inverse probability weights. The average causal effect is estimated via:
$$\hat{\psi}_{IPW} = n^{-1} \sum_i \frac{Y_i X_i}{\widehat{\pi}(Z_i)} - n^{-1} \sum_i \frac{Y_i (1-X_i)}{1 - \widehat{\pi}(Z_i)}$$
Confidence intervals can similarly be calculated using a bootstrapping procedure. Instead, we used a robust variance estimator, which ignores the estimation of the nuisance model. Confidence intervals constructed from the robust variance result in conservative variance estimates for the ACE, with coverage expected to be at least 95\% when the nuisance model is properly specified.

\subsubsection*{Augmented inverse probability weighted estimator}
The AIPW estimator consists of the treatment and outcome nuisance model predictions. These predictions are combined via the following formula to estimate the average causal effect:
\begin{equation}
\begin{split}
\hat{\psi}_{AIPW} = & n^{-1} \sum_i \frac{Y_i X_i}{\widehat{\pi}(Z_i)} + \frac{\widehat{m_1}(Z_i) (\widehat{\pi}(Z_i) - X_i)}{\widehat{\pi}(Z_i)}\\
& - n^{-1} \sum_i \frac{Y_i (1-X_i)}{1 - \widehat{\pi}(Z_i)} + \frac{\widehat{m_0}(Z_i) (X_i - \widehat{\pi}(Z_i))}{1 - \widehat{\pi}(Z_i)}
\end{split}
\end{equation}
As indicated by their name, the AIPW estimator can be seen to include the IPW and an augmentation term. For inference, Wald-type confidence intervals were constructed from the following variance empirical influence curve-based estimator
\begin{equation}
\nonumber
\begin{split}
\widehat{Var}(\hat{\psi}_{AIPW}) = & \frac{1}{n-1} \sum_i \left( \left( \frac{Y_i X_i}{\widehat{\pi}(Z_i)} + \frac{\widehat{m_1}(Z_i) (\widehat{\pi}(Z_i) - X_i)}{\widehat{\pi}(Z_i)} \right) \right. \\
& - \left. \left(  \frac{Y_i (1-X_i)}{1-\widehat{\pi}(Z_i)} + \frac{\widehat{m_0}(Z_i) (X_i - \widehat{\pi}(Z_i))}{1 - \widehat{\pi}(Z_i)}\right) - \hat{\psi}_{AIPW} \right)^2
\end{split}
\end{equation}
This estimated variance is directly used to calculate Wald-type confidence intervals via
$$\hat{\psi}_{AIPW} \pm 1.96 \sqrt{\frac{\widehat{Var}(\hat{\psi}_{AIPW})}{n}}$$

\subsubsection*{Targeted maximum likelihood estimation}
TMLE similar consists of the treatment and outcome nuisance models. For ease of later notation, we define the so-called clever covariate as:
$$\widehat{H}(X_i, Z_i) = \frac{X_i}{\widehat{\pi}(Z_i)} - \frac{1 - X_i}{1-\widehat{\pi}(Z_i)}$$
Using the clever covariate, we target (or update) the predictions from the outcome nuisance model. The targeting step is done by first estimating $\epsilon$ in the following parametric working model
$$\text{logit}(Y_i) = \text{logit}(\widehat{m_X}(Z_i)) + \epsilon \; \widehat{H}(X_i, Z_i)$$
The estimate $\hat{\epsilon}$ is then used to update the untargeted estimates via the following formulas:
$$\widehat{m_1^*}(\widehat{m_1}, \widehat{H}(1, Z_i), \hat{\epsilon}) = \text{expit}\left(\text{logit}(\widehat{m_1}(Z_i) + \hat{\epsilon} \; \widehat{H}(1, Z_i))\right)$$
$$\widehat{m_0^*}(\widehat{m_0}, \widehat{H}(0, Z_i), \hat{\epsilon}) = \text{expit}\left(\text{logit}(\widehat{m_0}(Z_i) + \hat{\epsilon} \; \widehat{H}(0, Z_i))\right)$$
Therefore, we can use the targeted model predictions to calculate the average causal effect as:
$$\hat{\psi}_{TMLE} = \frac{1}{n}\sum_i \widehat{m_1^*}(\widehat{m_1}, \widehat{H}(1, Z_i), \hat{\epsilon}) - \frac{1}{n}\sum_i \widehat{m_0^*}(\widehat{m_0}, \widehat{H}(0, Z_i), \hat{\epsilon})$$
For inference, Wald-type confidence intervals were constructed from the following variance estimator:
\begin{equation}
\nonumber
\begin{split}
\widehat{Var}(\hat{\psi}_{TMLE}) = & \frac{1}{n-1} \sum_i \left( \left( \frac{Y_i X_i}{\widehat{\pi}(Z_i)} + \frac{\widehat{m_1^*}(\widehat{m_1}, \widehat{H}(1, Z_i), \hat{\epsilon}) (\widehat{\pi}(Z_i) - X_i)}{\widehat{\pi}(Z_i)} \right) \right. \\
& - \left. \left(  \frac{Y_i (1-X_i)}{1-\widehat{\pi}(Z_i)} + \frac{\widehat{m_0^*}(\widehat{m_0}, \widehat{H}(0, Z_i), \hat{\epsilon}) (X_i - \widehat{\pi}(Z_i))}{1 - \widehat{\pi}(Z_i)}\right) - \hat{\psi}_{TMLE} \right)^2
\end{split}
\end{equation}

\newpage
\subsection*{Section 3: Algorithm for double cross-fitting}
Below is a general algorithm for the double cross-fitting procedure.

Let $(O_1, O_2, ..., O_n)$ be an observed sample of independent and identically distributed observations, where $O_i=(X_i,Y_i,Z_i)\sim F$. For the following estimation problem, $X$ indicates the treatment / exposure of interest, $Y$ indicates the outcome of interest, and $Z$ is the sufficient adjustment set of confounders for the $X$-$Y$ relationship. 

For ease of notation, let the parameters for nuisance models be defined as $\gamma$ for the treatment model and $\eta$ for the outcome model, such that $\hat{\gamma}_k$ indicates the parameter estimated using data from split $k$. 

\begin{legal}
	\item \textbf{REPEAT} for $p$ different partitionings:
	\begin{legal}
		\item Randomly partition data into 3 approximately-equal sized data sets. The partitioned data sets should be non-overlapping, such that individual i only contributes to one of the partitions. Let the particular splits under the partition be indicated by $s=\{1, 2, 3\}$.
		\item \textbf{FOR} each of the 3 splits:
		\begin{legal}
			\item \textbf{ESTIMATE} the treatment nuisance model for $\Pr(X=1|Z=z;\gamma_s)$.
			\item \textbf{ESTIMATE} the outcome nuisance model for $E[Y|X=1,Z=z;\eta_s]$.
		\end{legal}
		\item \textbf{FOR} splits $s=1$:
		\begin{legal}
			\item \textbf{CALCULATE} predicted probabilities of treatment using the model from split $s=2$, $\Pr(X_i=1|Z_i;\hat{\gamma}_2)$.
			\item \textbf{CALCULATE} predicted values of the outcome under each level of exposure using the model from split $s=3$, $\hat{E}[Y_i|X=x,Z_i;\hat{\eta}_3]$.
			\item \textbf{CALCULATE} the pseudo-outcomes using the chosen doubly-robust method. For the AIPW estimator this implies the estimator described in Appendix Section 2. \\
			For TMLE this requires the targeting step and generating the targeted predictions described in Appendix Section 2.
			\item \textbf{CALCULATE} the ACE for the split $s=1$ by taking the mean difference in the pseudo-outcomes.
			\item \textbf{CALCULATE} the variance for the split $s=1$ via the relevant influence curve formula for the estimator.
		\end{legal}
		\item \textbf{REPEAT} step 1.3 for split $s=2$ with $\Pr(X_i=1|Z_i;\hat{\gamma}_3)$ and $\hat{E}[Y_i|X=x,Z_i;\hat{\eta}_1]$
		\item \textbf{REPEAT} step 1.3 for split $s=3$ with $\Pr(X_i=1|Z_i;\hat{\gamma}_1)$ and $\hat{E}[Y_i|X=x,Z_i;\hat{\eta}_2]$
		\item \textbf{CALCULATE} $\widehat{ACE}_p$ as the mean of the point estimates for all three splits
		\item \textbf{CALCULATE} $\widehat{Var}(\widehat{ACE}_p)$ as the mean of the variances for all three splits 
	\end{legal}
	\item \textbf{CALCULATE} the median of the point estimates from the $p$ different partitions as the 
	\[\widetilde{ACE} = \text{median}(\widehat{ACE}_p)\]
	\item \textbf{CALCULATE} the variance as the median of the variance from $p$ different partitions and the variance of $p$-specific point estimates from the overall point estimate
	\[\widehat{Var}(\widetilde{ACE}) = \text{median}(\widehat{Var}(\widehat{ACE}_p) + (\widehat{ACE}_p-\widetilde{ACE})^2)\]
\end{legal}
The number of splits in the above algorithm can be generalized to numbers greater than three. Additionally, the mean can be used instead of the median to calculate the overall point estimate and variance.

\subsection*{Section 4: Number of different partitions for the double cross-fitting procedure}
Using the same data set described in the context of a single simulated data set, we compared differing number of partitions. We explored 5, 10, 25, 50, 75, and 100 as different possible number of different partitions for the cross-fit procedure before calculating the overall point estimate. Each of these partitions was re-ran on the data 100 times and the point estimates were extracted. Therefore, our plot of the number of partitions can be viewed as the stability of the point estimate under re-running the estimator with p different partitions. In the ideal performance, the point estimate would be invariant to re-running the double cross-fit procedure on the same data set with different possible partitions. Point estimates across different partitions for both DC-AIPW and DC-TMLE were further paired with correctly specified parametric models and super-learner for nuisance model estimation. 

\begin{figure}
	\centering
	\caption {Variation in point estimates for 100 re-runs of the double cross-fit algorithm by different number of partitions}
	\includegraphics[width=0.9\linewidth]{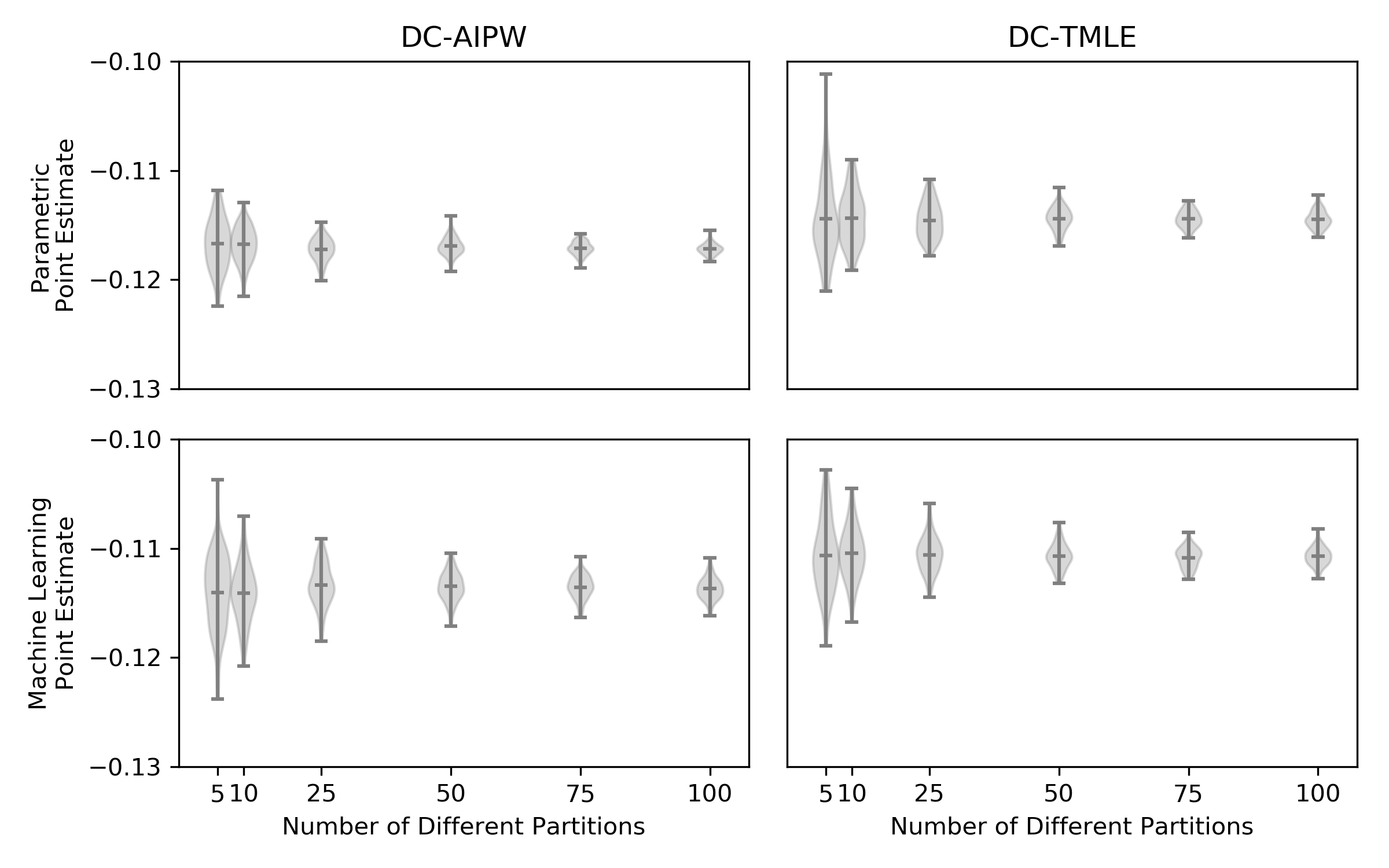}
	\label{afig}
	\floatfoot{DC-AIPW: double cross-fit augmented inverse probability weighted estimator, DC-TMLE: double cross-fit targeted maximum likelihood estimator. \\
	Violin plots are based on 100 re-runs of the double cross-fit procedure on the same data set.}
\end{figure}

As seen in the figure, fewer number of different partitions (e.g. 5, 10, 25) resulted in varying point estimates when re-ran on the same data set. At higher number of different partitions (e.g. 75, 100) there was less variation between re-runs of the double cross-fit procedure. These results suggest a large number of different partitions should be used in practice.

\end{document}